\documentclass[aps,amsmath,amssymb,twocolumn,prl,longbibliography]{revtex4-1}
\usepackage{graphicx} 
\usepackage{amsfonts,amsmath,amssymb}
\usepackage{amsthm}
\usepackage{dcolumn}
\usepackage{dsfont,bm}
\usepackage[colorlinks=true,linkcolor=blue,pagecolor=blue,filecolor=blue,menucolor=blue,urlcolor=blue,citecolor=blue,anchorcolor=blue]{hyperref}
\usepackage{xcolor}
\usepackage{soul} 
\usepackage{amsbsy}
\usepackage{float}

\usepackage{bm}

\begin{document}
\title{Movable Dirac Points with Ferroelectrics: Kink States and Berry Curvature Dipoles}
\author{Konstantin S. Denisov$^{1}$}
\author{Yuntian Liu$^{1}$}
\author{Igor {\v{Z}}uti{\'c}$^{1}$}
\affiliation{$^{1}$Department of Physics, University at Buffalo, State University of New York, Buffalo, NY 14260, USA}

\begin{abstract} 
Two-dimensional (2D) Dirac states and Dirac points with linear dispersion are the hallmark of graphene, topological insulators, semimetals, and superconductors. Lowering a symmetry by the ferroelectric polarization opens the gap in Dirac points and introduces finite Berry curvature. Combining this with Dirac points detached from high symmetry points of the Brillouin zone offers additional ways to tailor topological properties. We explore this concept by studying topological phenomena emerging in 2D materials with movable Dirac points and broken out-of-plane mirror reflection. The resulting topological kink states and Berry curvature dipoles are changed by movable 2D Dirac points with experimental signatures in electrical conductance and 
second-harmonic  nonlinear Hall conductivity. We identify materials platforms where our predictions can be realized and support that with the
first-principles results for  Cl$_2$Rh$_2$S$_2$-GeS junction.
\end{abstract}
\date{\today}
\maketitle

Ferroelectrics (FEs) with their switchable and nonvolatile polarization have been studied for decades in logic and memory applications~\cite{Mikolajick2021:JAP} 
as well to transform materials by extending the range of available electrostatic doping~\cite{Ahn2006:RMP}. With the recent discovery of two-dimensional (2D) FEs 
and their integration in many van der Waals (vdW) heterostructures, FE control has been further expanded to strongly modulate optical, transport, and magnetic properties 
as well as to manipulate topological states~\cite{Marrazzo2022:npj2D,Wang2017:2DM,Zhang2020:NL,Huang2022:NL}, a long-standing problem for realizing 
electrically controllable topological materials.

A desirable control over topological properties in 2D can be established by manipulating
{Dirac or Weyl} electrons~\cite{Armitage2018:RMP,Wan2011:PRB}, appearing when valence and conduction band approach
each other at {Dirac (Weyl)} points in the Brillouin zone (BZ) to form a linear dispersion. 
{Unlike in the 3D case, the protection of these gapless points in 2D is not robust and requires the presence of nonsymmorphic 
lattice symmetries~\cite{Young2015:PRL,Jin2020:PRL,Wieder2018:S,Herrera2023:CP,Liu2022:SR}. In absence of the protection, 
the gap opening  favors topological effects by realizing sub-bandgap edge states and equips Dirac states with finite Berry curvature,} 
$\Omega_{\bm{k}}^{nm}$, where $\bm{k}$ is the wave vector, that provides a local gauge-invariant information on geometrical structure of electron wave function 
in $k$-space between $n$ and $m$ bands~\cite{Vanderbilt:2018}. The integral of $\Omega_{\bm{k}}^{nm}$ over the entire BZ is a quantized value, a topological index, 
describing global, topological structure of electronic bands~\cite{Vanderbilt:2018,Xiao2010:RMP}. 
{The gap opening of Dirac points can be induced by spin-orbit coupling (SOC), 
usually switching a system to the $Z_2$ spin Hall insulator~\cite{Kane2005:PRL,Kane2005:PRLb}, 
or by breaking point-group symmetries, with the formation of electronic kink
states~\cite{Martin2008:PRL,Semenoff2008:PRL,Jung2011:PRB,Zhou2021:PRL,Ju2015:N,Li2018:S,Huang2024:S}.} 

{The most studied 2D Dirac material,  
graphene, has its Dirac states located at fixed and high-symmetry points ($K/K'$) in the BZ~\cite{CastroNeto2009:RMP}. 
Movable Dirac points in 2D systems, not constrained to such high-symmetry points, are expected to support tunable topological phenomena. 
After the first demonstration 
in interface-doped 2D phosphorene~\cite{Kim2015:S,Kim2017:PRL,Ehlen2018:PRB,Jung2020:NM}, 
there is a growing number of other 2D materials with movable Dirac~\cite{Lu2016:npjCM,Lu2022:NC,Liu2023:PSS,Kumar2023:JPCC} and Weyl points~\cite{Lu2024:NC,Liu2023:NL}. Combining symmetry-breaking induced $\Omega_{\bm k}$ with an extra degree of freedom in the position of Dirac points is an unexplored playground for tunable topological effects.}

\begin{figure}[t]
\includegraphics[width=0.98\columnwidth]{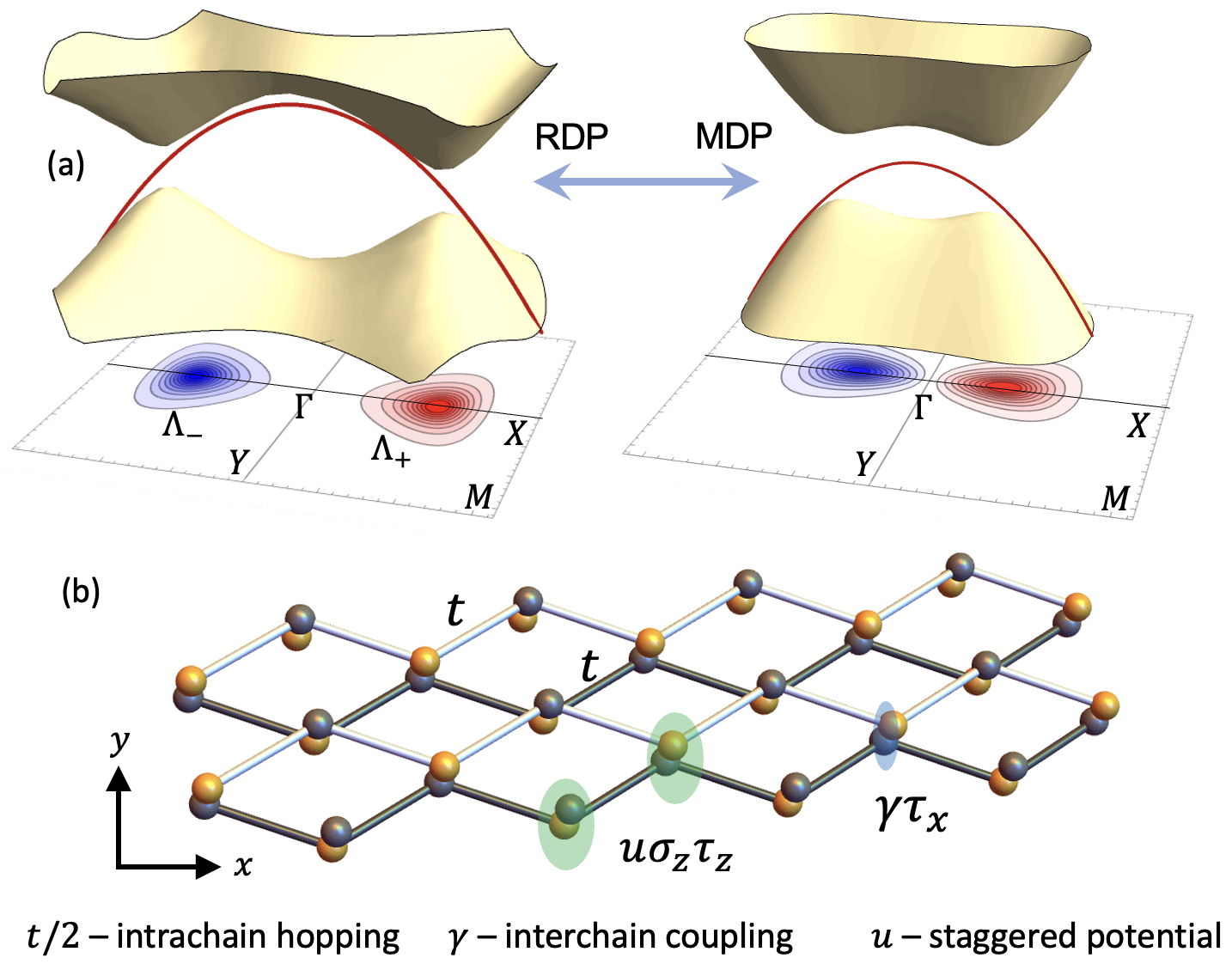} 
\caption{{(a)} Movable gapped Dirac points have a valley-dependent ($\Lambda_\pm$) and sign-altering Berry curvature across the Brilluion zone, shown in color. Left/Right: Remote/merged Dirac points (RDP/MDP). The energy dispersion of the resulting topological kink states (red line) is transformed between RDP/MDP.
$\Gamma$, X, Y, M, are the symmetry points. 
{(b) Our  lattice model in the low-symmetry configuration  hosting gapped moveble Dirac points, $\sigma$ ($\tau$) are Pauli matrices for the two 
inequivalent sublattices (different chains).}} 
\label{fig:1}
\end{figure}


In this work we study topological phenomena emerging in 2D materials with movable Dirac points that can be further controlled by FEs. The in-plane {FE polarization} 
realizes a sign-altering dipole configuration of {$\Omega_{\bm{k}}^{nm}$}, 
{and moving/fusing of Dirac points, as illustrated in Fig.~\ref{fig:1}, e.g. by lattice deformations,} alter $\Omega_{\bm{k}}^{nm}$-driven effects. Specifically, we focus on transforming topology-enforced kink electronic states 
at structural domain walls (DWs) {separating opposite FE polarizations.} 
Movable Dirac points 
control the measured conductance from kink states sensitive to the crystallographic orientation of DWs, 
and they also 
modify the dipole character of $\Omega_{\bm{k}}^{nm}$, which is responsible for switching between the zero and nonzero SHG in the nonlinear Hall conductivity. 
Our analysis uses
a transparent tight-binding (TB) model in 
Fig.~\ref{fig:1}(b) and first-principles methods for 2D materials with movable Dirac points, Supplemental Material (SM)~\cite{SMKS}.

{A prototype 2D model hosting movable Dirac points has} 
two coupled zigzag chains,  
{each with two inequivalent sublattices, shown in Fig.~\ref{fig:1}. } 
{A single chain represents  Su-Schrieffer-Heeger model~\cite{Su1979:PRL}, an early  example of a topological system.} 
{With a}  finite shift between the chains along $\hat{\bm{z}}$ the lattice
consistent with the {2D phosphorene}~\cite{Akhtar2017:npj}.
A symmetric configuration 
of the model  {without a staggered potential [$u=0$, Fig.~\ref{fig:1}(b)]} 
has the  point group symmetry, $\mathcal{D}_{2h}$, 
{and the spatial inversion, $\mathcal{P}$.}
The model is 
described by 
the {nearest-neighbor} 
hopping, 
$t/2$, equal for each chain, and
the coupling between two sites from different chains, $\gamma$~\cite{SMKS}. 
The {effective} {$k$-space spinless $4\times4$} Hamiltonian {(2 chains and  
sublattices)} is 
\begin{equation}
	\label{eq:Ham1}
	H_0 = - t  ( h_x \sigma_x + h_y  \sigma_y \tau_z) - \gamma  \tau_x,
\end{equation}
where $\sigma, \tau$ 
are Pauli matrices related to inequivalent sublattices
and different chains, respectively. 
The complex function $h = h_x + i h_y $, with $h = \cos{ \left(k_x a_0/2\right) } e^{i k_y b_0/2}$, 
accounts for the geometrical structure of the tight-binding lattice, 
$a_0, b_0$  are the lattice parameters along $x,y$ axes. 

In the basis of anti(symmetric) wave function combinations with respect to the chain index, 
$\psi_{s,\nu} = (\varphi_\nu ,  \sigma_x \varphi_\nu )^T/\sqrt{2}$ 
and $\psi_{a,\nu} = (\varphi_\nu , - \sigma_x \varphi_\nu )^T/\sqrt{2}$, where
$\nu = \pm$,  with $\varphi_+ = (1,0)^T$ and  $\varphi_- = (0,1)^T$, 
{$H_0$ now} splits into two independent $2\times 2$ blocks 
$H_{\rm s,a}$ 
\begin{equation}
	H_{\rm s,a} = -  \left(t  h_x \pm \gamma \right) \sigma_x  - t  h_y  \sigma_y, 
\end{equation}
with energies $E_\nu = \nu  | t h  \pm \gamma | $. 
The doublet {$\psi_{\nu}^a$} has  two gapless Dirac points 
along the $\Gamma X$ line in BZ, located at $k_D^\pm$, where $\cos{(k_D^\pm a_0/2)}  =  \gamma/t$. 
We can understand the evolution from 
remote (RDP) to merged (MDP) Dirac points, 
depicted in Fig.~\ref{fig:1}, by changing  $\gamma/t$. 
At $\gamma/t = 0$, the two identical chains are decoupled. At $0<\gamma/t \ll 1$, there is an energy shift and a band crossing near the X-point, 
$k_X=\pi/a_0$.
As $\gamma/t$ increases, $k_D^\pm$ moves along the $\Gamma X$ line, 
away from $\pi/a_0$, with two RDPs hosting anisotropic Dirac electrons of opposite chirality 
in $\Lambda_{\pm}$-valleys for $0.3 \lesssim \gamma/t \lesssim 0.8$. 
With  $\gamma/t \rightarrow 1$
MDPs form, resulting in a complex, anisotropic band structure at the $\Gamma$-point and the gap opening at $\gamma/t \gtrsim 1$. 
 
Gapless Dirac points of $H_{\rm a}$ are protected independently by $\mathcal{P}$ and twofold rotations along $\hat{\bm{x}}, \hat{\bm{y}}$ axes~\cite{Lu2016:npjCM}. 
{With SOC usually transforming a system to the $Z_2$-topological phase~\cite{Lu2015:NL,Lu2016:npjCM,Jin2021:NL}, here we consider only} 
orbital mechanisms of gapping Dirac points, 
as in the low-symmetry lattice configuration from  Fig.~\ref{fig:1}(b), where
the binary filling of the lattice sites is modelled by the on-site, chain-dependent staggered potential $V = u \sigma_z \tau_z$. 
{For  $H=H_0+V$, a point group symmetry is reduced to $\mathcal{C}_{2v}$ lacking $\mathcal{P}$ but respecting the mirror reflection plane, $\mathcal{M}_x$, compatible with {$\hat{\bm{y}}$} FE polarization.}
With $\eta \equiv {\rm sign}(u) = \pm 1$, we differentiate two inequivalent lattice configurations related by $\mathcal{P}$
and associate it with the direction of the spontaneous FE polarization. 

Our considered $H$ remains block diagonal in the $\psi_{s,a}^{1,2}$ basis, while
the reduced $2\times2$ block part is
\begin{eqnarray}
	\label{eq:Hred}
	H_{\rm a} = -  (t h_x - \gamma) \sigma_x - t h_y \sigma_y  + u \sigma_z,
\end{eqnarray}
with the spectrum, $E = \pm \sqrt{|t h-\gamma|^2 + u^2}$, 
acquiring finite energy gap $E_{g} = 2u$ at $\gamma < t$. 
The eigenfunctions  are
$	\varphi_k^+  =  (a_1, a_2 e^{i \theta_k})^T / \sqrt{2} $, 
$	\varphi_k^-  =  (a_2, - a_1 e^{i \theta_k})^T / \sqrt{2} $, 
where $a_{1,2} = \sqrt{1 \pm u/E}$ and $\theta_k = {\rm Arg}[\gamma - t h]$. The Berry curvature across the 
BZ is $\Omega_{\bm{k}}
=\Omega_{\bm{k}}^{cc} =\Omega_{\bm{k}}^{vv}=\Omega_{\bm{k}}^{cv}$, with
\begin{equation}
	\label{eq:Omega}
	\Omega_{\bm{k}} = (u t^2/16 E^3) a_0 b_0 \sin{\left( k_x a_0 \right)}, 
\end{equation}
where c (v) denote the conduction (valence) band. 
$\Omega_{\bm k}$ has the dipole configuration with  
$\pm$ peaks at $k_D^\pm$
[Fig.~\ref{fig:1}],  swapped  
for $\eta \rightarrow -\eta$ and realizing 
the valley-FE coupling~\cite{Zheng2022:JAP} for RDP at $\Lambda_{\pm}$. 

For the RDP, the integral of the Berry field 
in the valence band, $\Omega_{\bm{k}}^{vv} = \Omega_{\bm{k}}$, 
gives the Berry flux of opposite sign 
$Q_\pm = \pm \eta /2$ for $\Lambda_{\pm}$,
which ensures zero overall topological charge, 
$Q_+ + Q_- = 0$, and  
no quantum spin Hall states~\cite{Kane2005:PRL}. 
{From the index theorem and valley Chern numbers~\cite{Shen:2012,Qiao2011:PRL,Zhang2013:PNAS}}, 
following the 
Volkov-Pankratov model~\cite{Volkov1985:SJETPL}, 
gapped Dirac points of Eq.~(\ref{eq:Hred}) 
still lead to the formation of topological states. 
{The corresponding} kink states are along structural DWs separating 
regions with inverted mass terms and {yield} counterpropagating 1D channels with valley-locked direction of the velocity, 
as in  hexagonal lattices, including graphene~\cite{Martin2008:PRL,Semenoff2008:PRL,Jung2011:PRB,Ju2015:N,Li2018:S,Zhou2021:PRL}.

In our case, the kink states emerge at the DWs between regions 
separating opposite FE polarizations, $\eta=\pm 1$, 
and can be further combined with the quantum spin Hall states from the topological boundaries 
to suppress disorder-induced scattering~\cite{Zhou2021:PRL}.  
In Fig.~\ref{fig:2} we show how the movable {and anisotropic}
Dirac points lead to  \noindent{the} kink states for DWs parallel to x (DWx) and y axis (DWy), described in (a)-(e) and (f)-(i), respectively. We evaluate~\cite{Groth2014:NJP} 
the tight-binding lattice model of $H$ for a slab geometry of a finite width $W$ along y (x) for DWx (DWy). The DW profile is introduced by taking $u$ to have opposite signs at the boundaries $\pm W/2$.

For DWx, {$k_x$ is as a good quantum number, and} the topological kink states with 
valley-velocity locking are formed in Fig.~\ref{fig:2}(a) for RDP. 
The dispersion $E(k_x)$, shown {(red lines)} in Fig.~\ref{fig:2}(a), 
reflects the presence the kink states crossing the energy bandgap that are coupled to $\Lambda_{\pm}$ valleys. 
The corresponding spatial structure of the probability density (summed over chain and sublattice indices, $\gamma/t=1/2$) for a subgap state, $|E| < u$, is given in Fig.~\ref{fig:2}(c) together 
with the site-dependent $u(y)$ (dashed line) indicating the kink state pinned to DWx. If we turn to DWy, in Fig.~\ref{fig:2}(f) we show the dispersion 

\begin{widetext}
\begin{minipage}{\linewidth}
\begin{figure}[H]
\flushleft			
\includegraphics[width=0.99\textwidth]{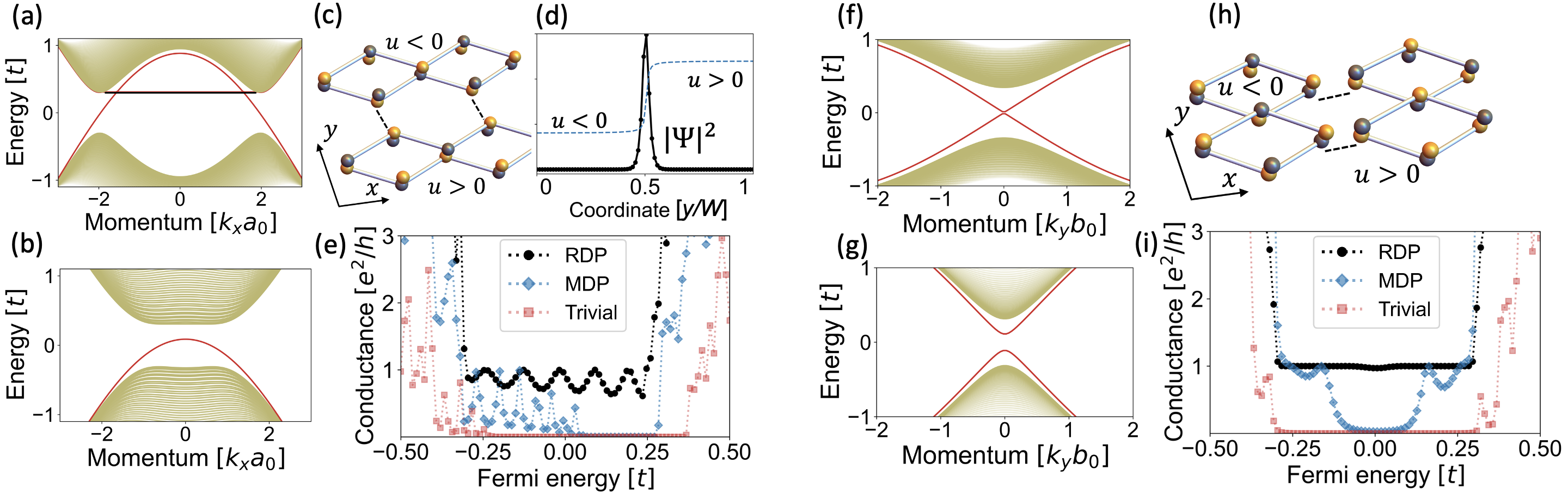}
\caption{Topological kink states at DWs, parallel to $x$ (DWx) and $y$ axis (DWy) for remote {($\gamma/t=0.6$)} and merged {($\gamma/t=0.95$)} Dirac points with $u=\pm 0.3t$. (a) [(b)] RDP (MDP) for DWx, the red color: the kink state dispersion. (c) The slab geometry with DW 
	($\pm u$ the two boundaries, $y=\pm W/2$). 
(d) DWx profile with kink state probability density. (e) The conductance $G(\mu)$ as a function of the Fermi energy, $\mu$. (f) [(g)] RDP (MDP) for DWy. (h) The slab geometry.  (i) The corresponding $G(\mu)$.}
\label{fig:2}
\end{figure}    
\end{minipage}
\end{widetext}
$E(k_y)$ from the slab geometry of Fig.~\ref{fig:2}(h) for RDP, {where the   two kink states with opposite velocity from $\Lambda_{\pm}$-valleys intersect at $k_y=0$.}

Movable Dirac points imply a possibility to realize their fusion,
accompanied by the dissolution of the kink states, a process that reveals itself in the change of the  
two-point conductance, $G(\mu)$, as a function of the Fermi energy, $\mu$. 
Remarkably, the dissolution trajectory of the kink states depends on the DW orientation. 

{For DWx,  moving from RDP [Fig.~\ref{fig:2}(a)], 
with kink states crossing the entire bandgap 
to MDP [Fig.~\ref{fig:2}(b)], 
pushes the kink states away from the conduction band to merge with the valence band after the fusion, at $\gamma/t > 1$, 
which affects $G(\mu)$ [Fig.~\ref{fig:2}(e)].}
For RDP,  $G \lesssim e^2/h$  for all energies inside the bulk bandgap, indicating 
the presence of one open channel (per spin).  For MDP, $G \rightarrow 0$ when 
$\mu \rightarrow u$, as expected from the absence of open ballistic channels in this energy range, seen in Fig.~\ref{fig:2}(b). 
In the trivial regime, $G=0$ inside the bandgap. 

A different dissolution is realized for DWy: The transformation from RDP to MDP in Figs.~\ref{fig:2}(f) and \ref{fig:2}(g) is realized by the gap opening in the spectrum of the kink states 
{due to mixing of $\Lambda_{\pm}$ valleys} and the subsequent merging of the edge states with the conduction and valence bands, separately, at $\gamma/t > 1$. 
The calculated $G(\mu)$ in Fig.~\ref{fig:2}(j) shows these changes for  RDP, MDP, and trivial regimes. In contrast to the RDP signature of
$G = e^2/h$ for a large range of $\mu$,  for MDP $G=0$ in the middle of the bandgap, as expected from Fig.~\ref{fig:2}(g). 
Taken together, the conductance offers an important experimental fingerprint for the predicted the kink states.

{To demonstrate the universality of our description for topological kink states based on the 
movable Dirac points, we screened topological semimetals 
without SOC from the Topological 2D Materials Database~\cite{Petralanda2024:X,Jiang2024:X} and identified 
36 materials in Table S1~\cite{SMKS} with clean Dirac cones on high-symmetry lines near the Fermi level. 
The layer group symbols and the positions of the Dirac points for each material are also included, covering both square and hexagonal systems.
The movable Dirac point pairs can vary from RDP to MDP near $\Gamma, X, Y, M$ or $K$ points.}

\begin{figure}[t]
	\centering
	\includegraphics[width=.5\textwidth]{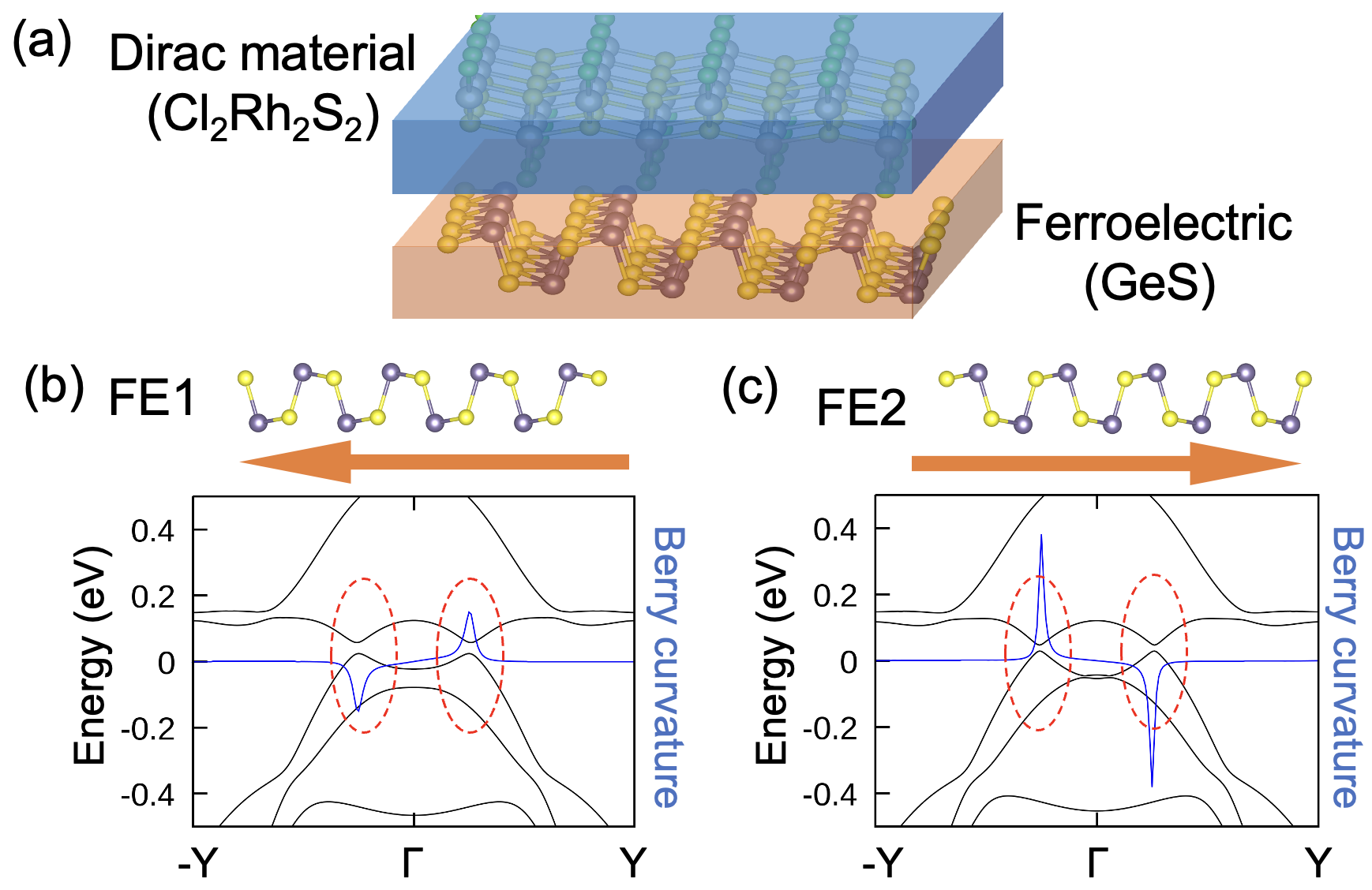}
	\caption{(a) Schematic diagram and structure of a
	junction composed of  a Dirac material Cl$_2$Rh$_2$S$_2$ with  movable Dirac points and ferroelectric GeS. 
	(c), (d) Band structure (black lines) and Berry curvature (blue line) of the Cl$_2$Rh$_2$S$_2$-GeS junction in (c) the FE1 state and (d) the FE2 state, 
	where the orange arrows denote the FE polarization.}
	\label{fig:3}
\end{figure}

{While one realization of the FE-controlled kink states implies 2D FEs with an in-plane polarization, 
another one could be through proximity effects~\cite{Zutic2019:MT} from FE substrates.
We confirm this other platform in Fig.~\ref{fig:3}  from first-principles calculations for a junction of monolayers of the Dirac material Cl$_2$Rh$_2$S$_2$ 
and the group-IV monochalcogenide ferroelectric GeS~\cite{Barraza2021:RMP}. 
Cl$_2$Rh$_2$S$_2$ has a rectangular crystal structure with the Cl-Rh-S chains oriented along the a axis and alternately arranged
along the b axis~\cite{SMKS}, resembling 
the structure of our TB model. By applying an  in-plane polarization along the a axis of  
the GeS substrate [Figs.~S1(a), S1(b)],  
the paired Dirac points along the $\Gamma-Y$ high-symmetry line open an 
energy gap of $E_g=33.3$ meV [Fig.~\ref{fig:3}(b)] and $E_g=18.0$ meV [Fig.~\ref{fig:3}(d)]
and exhibit opposite Berry curvatures. This Berry curvature dipole is tunable by two inequivalent GeS configurations 
FE1  [Figs.~\ref{fig:3}(b)] 
and FE2 [Fig.~\ref{fig:3}(c)] with opposite FE polarization 
and robust under SOC [Figs.~S1(c)-1(f)].} 

\begin{figure}[t]
\centering
\includegraphics[width=.44\textwidth]{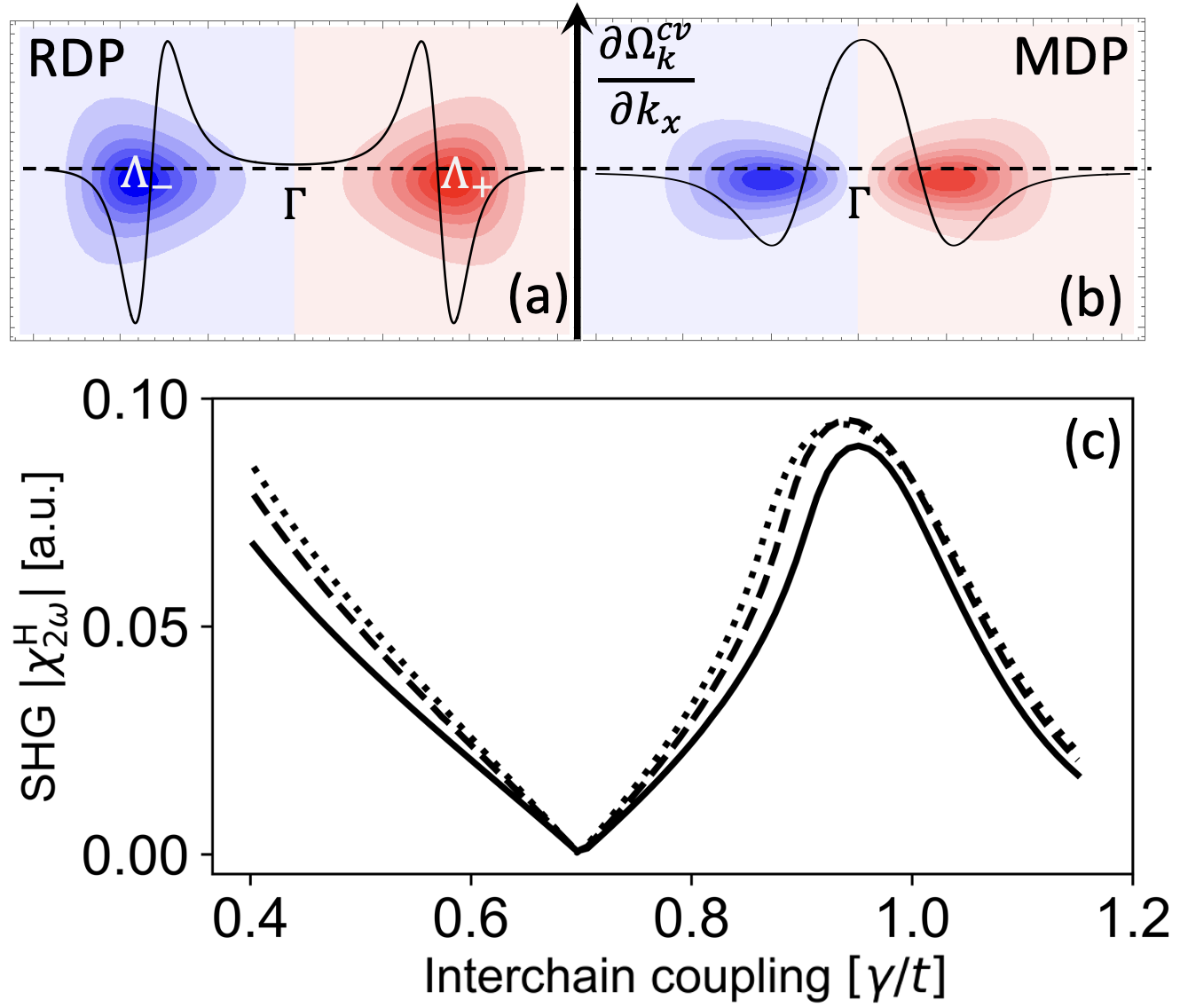} 
\caption{The evolution between the (a) RDP {($\gamma/t=0.6$)}
and (b) MDP {($\gamma/t=0.95$)} is accompanied by the change of the Berry curvature (color map in BZ)
and its dipole contribution  $\partial{\Omega^{\rm cv}_{\bm{k}}}/\partial{k_x}$ (solid lines). (c) The switching between zero and nonzero $|\chi_{2\omega}^{\rm H}|$ 
for subgap excitation with $ \delta \hbar \omega = 0.1,0.2,0.3 u$ (solid, dashed, and dotted lines) by changing the position of the Dirac points with $\gamma/t$.}
\label{fig:4}
\end{figure}

{In addition to the  dipole structure of $\Omega_{\bm k}$ for  the topological kink states, the in-plane FE polarization also induces 
 second-order electromagnetic responses, such as the nonlinear Hall effect (NHE)~\cite{Sodemann2015:PRL,Xu2018:NP,Ma2019:N,Zhang2022:PRB,Bandyopadhyay2023:MTE}.}	
Transforming the Berry curvature dipoles for NHE has been discussed for the linear photo-galvanic effect~\cite{Xu2018:NP,Du2018:PRL,Ma2019:N,Kang2019:NM,Orenstein2021:ARCMP,Zhang2022:PRB,Bandyopadhyay2023:MTE,Sinha2022:NP,Chakraborty2022:2DMat}
or the second harmonic generation (SHG)~\cite{Kim2019:NC,He2021:NC,Okyay2022:CP}.  
{Here, we focus on the imaginary part of the 
subgap SHG-Hall conductivity in the insulating regime, $ \chi_{2 \omega}^H \equiv {\rm Im}[\chi_{yxx}(-2 \omega, \omega, \omega)]$, 
and its previously unexplored evolution for different positions of Dirac points. We evaluate
$\chi_{2 \omega}^H$ in the clean limit and for the 
low-symmetry configuration of our lattice model from Eqs.~(\ref{eq:Hred}) and (\ref{eq:Omega}).}  
$\chi_{2 \omega}^H$ can be obtained 
in terms of the difference between {the} Berry connections~\cite{Aversa1995:PRB,Sipe2000:PRB} for 
occupied and unoccupied bands,  
$ \chi_{2 \omega}^H \propto \sum_k \omega_{\rm cv}^{-4} {\rm Im}[v^y_{cv} v^x_{vc} \delta v^x_{cv} ] (f_v-f_c) \delta(2\omega - \omega_{\rm cv})$, 
where $v^{\alpha}_{cv}$ is the matrix element of the velocity operator, 
$\delta v^x_{cv} = v_{cc}^x - v_{vv}^x$, 
$f_{c,v}$ are the distribution functions, while in the $\delta$-function $\hbar \omega_{\rm cv} = E_g$. 
For the isolated two-band model~\cite{Okyay2022:CP}, 
another approach is to use the contribution of the Berry curvature dipole,
$\chi_{2 \omega}^H \approx (\pi e^3/\hbar^2 \omega^2) D_{cv}^x$, with 
\begin{equation}
	{D}_{cv}^x = \sum_k \frac{\partial \Omega_{\bm{k}}^{cv}}{\partial k_x}
	\Theta(2\omega - \omega_{\rm cv}),
	\label{eq:Dipole}
\end{equation}
where $\Theta$ is the step function,
and $\Omega^{cv}_{\bm{k}}= \Omega_{\bm{k}}$ is the interband Berry curvature. 
With both approaches, we obtained matching results for $\chi_{2 \omega}^H$, summarized in Fig.~\ref{fig:4}.

{Switching $\chi_{2 \omega}^H$ between zero and nonzero values follows the transformation} 
of the band structure and is related to the profile of the Berry curvature dipole from Eq.~(5), 
shown for RDP and MDP in Figs.~\ref{fig:4}(a) and \ref{fig:4}(b). 
{In Fig.~\ref{fig:4}(c)} we present the evolution of $ \chi_{2 \omega}^H$ for a constant  {attenuation frequency}, $\delta \omega = 2\omega - \omega_{\rm cv}  \ll \omega_{\rm cv}$ when moving the position of Dirac points by changing $\gamma$ for different $\delta \omega$. 
For RDP at  $\gamma \approx 0.7 t$ and a largely decoupled valleys, $\chi_{2 \omega}^H \rightarrow 0$, as 
$\partial \Omega_{\bm{k}}^{\rm cv}/\partial k_x$ from Fig.~\ref{fig:4}(a) contains one symmetric oscillation for each valley. 
In contrast, for MDP at $\gamma \approx t$,  there is a highly anisotropic band structure at the $\Gamma$ point [Fig.~\ref{fig:1}], seen also in Fig.~\ref{fig:4}(b) {for $\gamma = 0.95 t$}, with a partial rectification of the Berry curvature dipole, disappearing quickly after the fusion, $\gamma > t$,  and leading to the peak in $\chi_{2\omega}^{\rm H}$ at $\gamma \approx t$, visible in Fig.~\ref{fig:4}(c). 
Remarkably, another enhanced  region of $\chi_{2 \omega}^H$ occurs for RDP by decreasing $\gamma$.
In the intermediate range $0.3 \lesssim \gamma/t \lesssim 0.6$, 
when the two Dirac points are far apart,  the magnitude of $\chi_{2\omega}^{\rm H}$ in Fig.~\ref{fig:4}(c)
is comparable with its peak value at $\gamma \approx t$ for the fused Dirac points. 
A finite SHG for isolated Dirac points was discussed for tilted Dirac cones~\cite{Du2018:PRL,Duan2023:NM,Liang2021:PE,Gao2021:OE}, 
trigonal warping as in graphene~\cite{Margulis2013:JPCM,Battilomo2019:PRL,Golub2014:PRB}, 
or oblique photon incidence~\cite{Mikhailov2008:JPCM,Dean2010:PRB,Glazov2014:PR}.


{For implementing movable Dirac points,} 
it would be important to consider electro-elastic coupling~\cite{Zhang2020:NL,Marrazzo2022:npj2D,Liang2021:npjCM} 
realizing the band structure changes 
(e.g. in TB parameters of
2D materials~\cite{Verberck2012:PRB,Naumis2017:RPP,Yan2013:NC})
in response to lattice deformations and FE polarization. 
This scenario is 
corroborated by early experiments~\cite{Kim2015:S,Kim2017:PRL,Lu2022:NC} and first-principles 
studies~\cite{Lu2016:npjCM}, showing that the position of the known unpinned Dirac points is strongly influenced by the strain and environment effects. 

{We verify that the Dirac points are movable by performing first-principles calculations of a deformed pristine monolayer Cl$_2$Rh$_2$S$_2$. 
Our results are summarized in~\cite{SMKS} and suggest 
that including positive (negative) deformation along $a$ axis results in 
the Dirac points moving away (towards) each other.  
While $+5\%$ of deformation along $a$-axis almost doubles the distance between the Dirac points, applying $-2.5\%$ results in their complete fusion at the $\Gamma$-point, with the subsequent opening of an energy gap~$\sim 40$~meV at $-5\%$ deformation. In terms of our TB model, the obtained transformation corresponds to the change in the parameter $\gamma/t$ from $\gamma/t = 0.7$ at $5\%$ to $\gamma/t = 1$ at $-2.5\%$ and $\gamma/t = 1.2$ at $-5\%$. The strain along $b$ axis also results in the motion of Dirac points, but with smaller magnitudes.}

With our work it is now possible to consider dynamical topological phenomena driven by time-oscillating positions of Dirac cones.
Having gapped movable Dirac points allows us to pump charges from bulk to the edge kink states without closing the energy gap, 
a precursor of intriguing dynamics not realized in SOC-induced topological spin Hall states~\cite{Kane2005:PRLb}. 
With a growing class of topological kink states~\cite{Huang2024:S}, 
FE control could allow probing their non-Abelian statistics.

{In contrast to the out-of-plane FE polarization~\cite{Guan2019:AEM} usually
considered for opening the gap in 2D Dirac states, we study movable Dirac points in systems with an in-plane polarization. Such in-plane FEs may also simplify the transfer methods for fabricating vdW heterostructures. Moreover,} 
the emergence of kink states only requires DWs in lattices with the broken inversion symmetry, even without FEs,  
and thus are supported 
in a larger class of 2D materials, including quasi-Dirac semimetals~\cite{Li2024:PRL}
and antiferromagnets~\cite{Liu2022:PRX}. 
{Our conclusions also pertain to 2D magnets 
with a single pair of Weyl points close to the Fermi level.
In addition to transport probes~\cite{Ju2015:N,Li2018:S}, the kink states could also be detected by the angle-resolved photoemission spectroscopy~\cite{Hofmann2021:ES}. 
We anticipate similar evolution of electronic kink states  
in 2D materials with Dirac points located at generic positions in  the BZ, 
unpinned from any high-symmetry lines of  the 
BZ~\cite{Lu2022:NC}.}
One can also envision that our analysis of the tunable topological kink states have their 
superconducting counterparts as the FE influence, trough the tunable proximity effects in 
vdW heterostructures~\cite{Zutic2019:MT}, could be transferred to the 
growing class of 2D superconductors~\cite{Amundsen2024:RMP}.
 
\acknowledgments
This work was supported by the Air Force Office of Scientific Research under Award No. FA9550-22-1-0349.  Computational resources were provided by the UB Center for Computational Research.

\bibliography{KS_Ref}
\end{document}